\documentclass{article}

\usepackage{arxiv}

\usepackage[utf8]{inputenc} 
\usepackage[T1]{fontenc}    
\usepackage{hyperref}       
\usepackage{url}            
\usepackage{booktabs}       
\usepackage{amsfonts}       
\usepackage{nicefrac}       
\usepackage{microtype}      
\usepackage{lipsum}
\usepackage{graphicx}
\graphicspath{ {./images/} }
\usepackage{amssymb}
\usepackage{amsmath}

\title{A Physical Model Approach to Order Lot Sizing}

\author{
 Tania Daiana Tobares \\
 Grupo de Investigación Sistemas Complejos (SiCo)\\
 Facultad Regional San Rafael \\
  Universidad Tecnol\'{o}gica Nacional\\
  Mendoza, CP 5600 \\
  \texttt{tanitobares@hotmail.com} \\
  \And
 Margarita Miguelina Mieras \\
 Grupo de Investigación Sistemas Complejos (SiCo)\\
 Facultad Regional San Rafael \\
  Universidad Tecnol\'{o}gica Nacional\\
  Mendoza, CP 5600 \\
  \texttt{mmieras@frsr.utn.edu.ar} \\
  \And
 Fabricio Orlando Sanchez Varretti \\
 Grupo de Investigación Sistemas Complejos (SiCo)\\
 Facultad Regional San Rafael \\
  Universidad Tecnol\'{o}gica Nacional\\
  Mendoza, CP 5600 \\
  \texttt{fsanchez@frsr.utn.edu.ar} \\
     \And
   José Luis Iguain\\
 Instituto de Investigaciones Físicas de Mar del Plata (IFIMAR) \\
 Facultad de Ciencias Exactas y Naturales\\
  Universidad Nacional de Mar del Plata\\
  CONICET\\
  Mar del Plata, CP 7600 \\
  \texttt{iguain@mdp.edu.ar} \\
  \And
   Antonio José Ramirez-Pastor \\
 Instituto de Física Aplicada (INFAP)\\
 Universidad Nacional de San Luis\\
  CONICET\\
  San Luis, CP 5700 \\
  \texttt{antorami@gmail.com} \\
}

\begin{document}
\maketitle
\begin{abstract}
The growing need for companies to reduce costs and maximize profits has led to an increased focus on logistics activities. Among these, inventory management plays a crucial role in minimizing organizational expenses by optimizing the storage and transportation of materials. In this context, this study introduces an optimization model for lot sizing problem based on a physical system approach. By establishing that the material supply problem is isomorphic to a one-dimensional mechanical system of point particles connected by elastic elements, we leverage this analogy to derive cost optimization conditions naturally and obtain an exact solution. This approach determines lot sizes that minimize the combined ordering and inventory holding costs in a significantly shorter time, eliminating the need for heuristic methods. The optimal lot sizes are defined in terms of the parameter $\gamma=2C_O/C_H$, which represents the relationship between the ordering cost per order $C_O$ and the holding cost per period for the material required in one period $C_H$. This parameter fully dictates the system’s behavior: when $\gamma \leq 1$, the optimal strategy is to place one order per period, whereas for $\gamma > 1$, the number of orders $N$ is reduced relative to the planning horizon $M$ ($N<M$). By formulating the total cost function in terms of the intensive variable $N/M$, we consolidate the entire optimization problem into a single function of $\gamma$. This eliminates the need for complex algorithms, enabling faster and more precise purchasing decisions. The proposed model was validated through a real-world case study and benchmarked against classical algorithms, demonstrating superior cost optimization and reduced execution time. These findings underscore the potential of this approach for improving material lot sizing strategies.	
\end{abstract}


\section{Introduccion}
The optimal lot size in Material Requirements Planning (MRP) is a fundamental concept in inventory management and business logistics. Determining the appropriate lot sizes has significant implications for operational costs, supply chain efficiency, and customer satisfaction. Lot sizing is considered one of the most critical and challenging problems in production planning \cite{Karimi2003}.

Classical approaches to lot-sizing have been extensively studied in the literature. One of the fundamental techniques in inventory management is the Economic Order Quantity (EOQ) model \cite{Harris,Wilson,Caliskan}. Originally developed by F. W. Harris \cite{Harris} and later refined by R. H. Wilson \cite{Wilson}, this model states that the optimal lot size is achieved when the total inventory holding cost equals the total ordering cost.
Over the years, several extensions of the EOQ model have been proposed. Among them, two notable variations are the Fixed Order Quantity (FOQ) model \cite{FOQ} and the Periodic Order Quantity (POQ) model \cite{POQ}. The FOQ model is a variant of the EOQ model that follows a replenishment policy in which a fixed quantity of product is ordered whenever inventory reaches the reorder level. FOQ is a fixed policy that may be based on considerations beyond cost optimization. This approach has been widely adopted in logistics and inventory management \cite{FOQ}. Similarly, the Lot-for-Lot (L4L) technique is a lot-sizing method that generates exactly the quantity needed to meet the plan \cite{POQ, Noori1997}. 
The POQ model, on the other hand, is designed to address lumpy demand, which fluctuates over the planning period. The EOQ method can still be used to calculate the POQ, where the order quantity is determined based on the period obtained by dividing the EOQ by the demand per period \cite{FOQ}.

Another variation of the EOQ model is the Part-Period Balancing (PPB) algorithm, a lot-sizing heuristic suitable for discrete and time-varying demand data. First introduced in 1968 \cite{DeMatteis}, the PPB algorithm was developed as a single-stage lot-sizing procedure for environments where replenishment opportunities occur at discrete points in time. The order quantities generated by the algorithm are “discrete” in the sense that each represents the sum of an integer number of consecutive period requirements. The algorithm is based on the key EOQ property that, for each batch formed, the holding cost equals the ordering cost. The fundamental premise of the PPB approach is that the characteristics of the EOQ solution in a continuous and constant demand environment can be preserved when applied to a setting where demand varies between periods and inventory depletion occurs in discrete steps rather than continuously.

Another classical model is the Wagner-Whitin (W\&W) algorithm, widely used in production scheduling and inventory management in environments with limited capacity and variable demand. This method aims to minimize total costs by considering production, storage, and inventory holding costs, as well as capacity constraints \cite{Wagner}. Additionally, heuristic methods such as the Silver-Meal (SM) algorithm are employed to determine lot sizes in scenarios with highly variable demand \cite{Silver}. However, in cases where there are periods of zero demand, this method does not yield optimal results.
In the 1990s, Akimoto et al. \cite{Akimoto} proposed a Mixed Integer Linear Programming (MILP) model that addressed drum level control and the optimal distribution of fuel gas in steelworks power plants. Over time, this MILP model was extended and generalized to study various systems within the framework of the lot-sizing problem \cite{Golmohamadi}.

Beyond the classical works discussed in the previous paragraphs, the lot-sizing problem remains a critical topic in inventory management research \cite{Shekarabi,Cardona2020,Pooya2020,Bonsa2023,Arampatzis2024}. Researchers such as Hoseini Shekarabi et al. \cite{Shekarabi} have focused on optimizing order lot sizes while considering stochastic constraints, determining both the optimal number of lots and their respective volumes. Cardona-Valdés et al. \cite{Cardona2020} address a lot-sizing problem incorporating capacity constraints, lead times, batch orders, and shortages. They formulate and solve the problem using a Mixed Integer Linear Programming (MILP) approach, as the lead time for each item is supplier-determined, making it challenging for companies to plan purchases empirically. Pooya et al. \cite{Pooya2020} present a systematic approach to minimizing total costs and reducing nervousness in Material Requirements Planning (MRP) systems. The authors evaluate four lot-sizing policies across different scenarios: Lot-for-Lot (L4L), Fixed Order Quantity (FOQ), Periodic Order Quantity (POQ), and Economic Order Quantity (EOQ). Basa Bonsa et al. \cite{Bonsa2023} explore the lot-sizing problem in conjunction with supplier selection and order allocation under quantity discounts. To address this problem, they employ an efficient genetic algorithm enhanced with problem-specific operators. Finally, Arampatzis et al. \cite{Arampatzis2024} propose a variant of the Wagner-Whitin (W\&W) algorithm by integrating complementary cost factors. This modification improves the algorithm’s applicability, broadening its usefulness beyond its conventional scope.

In a recent study by our group \cite{Tania2023}, the lot-sizing problem was approached through the concept of clustering, where a cluster is defined as a set of elements within the same batch of orders \cite{Yin}. The clustering concept is widely applied across various scientific fields \cite{Stauffer}. Clustering analysis involves examining the different ways in which components of a system are grouped \cite{Yin,Stauffer}. In the case of one-dimensional clustering, as observed in various natural phenomena, the system is defined as a set of elements with similar characteristics, separated by empty spaces on both ends \cite{Zhang,Sciortino,Kaever,SanchezVarretti}.

In Ref. \cite{Tania2023}, the proposed model enables the determination of optimal lot sizes at minimal cost by employing a mathematical framework applicable to all material requirement options. The study provides both analytical and graphical insights into system behavior when it is considered as the set of clusters necessary to meet raw material demands for a given production period.

In this study, we continue addressing the cost optimization problem in ordering and storage within the lot-sizing model. This time, we introduce an optimization approach based on a physical system analogy, establishing that the material supply problem is isomorphic to a one-dimensional mechanical system of point particles connected by elastic elements. This analogy enables the natural derivation of cost optimization conditions, leading to an exact solution without relying on heuristic methods.

The proposed model determines lot sizes that minimize the combined ordering and inventory holding costs while significantly reducing computational time. To validate its effectiveness, we applied the model to a real-world case study and benchmarked it against classical lot-sizing algorithms. The results demonstrate superior cost optimization and reduced execution time, underscoring the potential of this approach to improve material lot-sizing strategies in practical applications.

The paper is structured as follows. Section \ref{MBD} introduces the model and fundamental definitions. The results are presented in Section \ref{results}. First, in Subsection \ref{primerparte}, the predictions of the proposed model are analyzed from a theoretical perspective. Then, in Subsection \ref{realcase}, the theoretical framework is applied to the analysis of real supply data from a concentrated pulp processing plant. Finally, general conclusions are presented in Section \ref{conclu}.

\section{Model and basic definitions}

\label{MBD}

We study a single-item, $M$-period economic lot-sizing model. Let $N$ denote the number of orders placed over the $M$ periods, $C_O$ the fixed ordering cost per order, and $C_H$ the fixed holding cost per period for the quantity of material required for one period. The simplest strategy to meet the demand for the $M$ periods consists of ordering the required quantity for each period at the beginning of that period ($N=M$). In this case, the order is consumed within the same period, resulting in no holding cost and a total cost of $C_T=MC_O$.

Introducing inventory storage ($N<M$) opens up various lot-sizing strategies. Identifying the optimal strategy that minimizes total cost is particularly valuable. Each strategy begins by ordering the material required for $n_1$ periods. After 
$n_1$ periods, these supplies are depleted, and additional material for $n_2$     periods is ordered. This process repeats iteratively until the $M$ periods are covered. Therefore, each strategy can be represented by a set of $N$ integers  $\{ n_i \}$ ($i=1,\dots,N$), satisfying:
\begin{equation}
	\sum_{i=1}^N n_i = M.
	\label{ligadura}
\end{equation}

As an illustrative example, we now analyze in detail a system with a planning horizon of $M=5$ and two orders ($N=2$). The first order satisfies the demands for periods 1 and 2 ($n_1=2$), while the second order fulfills the requirements for periods 3, 4, and 5 ($n_2=3$). The total cost associated with this strategy is given by:
\begin{equation}
	C_T(N=2,\{n_1=2,n_2=3\})= 2 C_O+ C_H + C_H + 2C_H.
	\label{eqm5}
\end{equation}
On the right-hand side of equation (\ref{eqm5}), the first term, $2 C_O$, represents the total ordering cost. The second term, $C_H$, corresponds to the holding cost of the material ordered at the beginning of the first period, stored during that first period, and consumed in the second period. The third term, $C_H$, represents the holding cost of the material ordered at the beginning of the third period, stored during that period, and consumed in the fourth period. Finally, the fourth term, 
$2C_H$, accounts for the holding cost of the material ordered at the beginning of the third period, stored during both the third and fourth periods, and consumed in the fifth period.

In general, for a strategy involving $N$ orders and a set $\{ n_i \}$, the total cost is expressed as:
\begin{eqnarray}
	C_T(N,\{n_i\}) & = & N C_O+ C_H \sum_{i=1}^N\sum_{j=1}^{n_i-1} j \nonumber \\
	& = &  N C_O+ 	\frac{C_H}{2}\sum_{i=1}^N n_i (n_i-1).
	\label{ct1}
\end{eqnarray}
By using equation (\ref{ligadura}), the last equation (\ref{ct1}) can be written as:
\begin{equation}
	C_T(N,\{n_i\})  =  N C_O+ \frac{C_H}{2}\sum_{i=1}^N n_i^2 - \frac{MC_H}{2}.
	\label{ct2}
\end{equation}

\begin{figure}[!htb]
\begin{center}
	\includegraphics[width=0.65\linewidth]{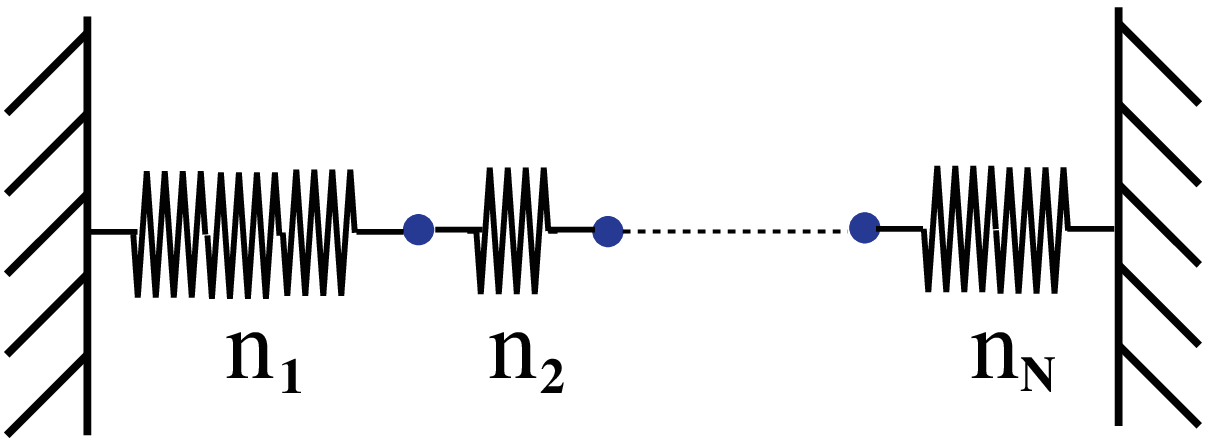}
	\caption{Schematic representation of an array of ($N-1$) particles (solid circles) connected by $N$ linear springs.}
	\label{fig:resortes}
\end{center}
\end{figure}

Note that equation (\ref{ct2}) corresponds, except for the additive constant
$ N C_O-MC_H/2$, to the potential energy of the one-dimensional system
shown schematically in Fig. \ref{fig:resortes}. That is, an array of ($N-1$) particles connected by $N$ linear springs of elastic constant $C_H$, with total length $ \sum_{i=1}^N n_i=M $.

Following this analogy, it is easy to see that, for a given $N$, the minimum of  $C_T(N,\{n_i\})=C_{T,0}(N)$ corresponds to the configuration in which all the springs have the same length, so that the force on every particle is zero, i.e.,
\begin{equation}
	n_i=\frac{M}{N}, \;\;\;\mbox{for}\;\; i=1,...,N.
	\label{min_config}
\end{equation}

This can be shown as follows. Let us write $C_T$ in terms of the $N-1$ independent variables $\{n_1,...,n_{N-1}\}$,
\begin{equation}
	C_T(N,\{n_i\})  =  N C_O+ \frac{C_H}{2} \left[ \sum_{i=1}^{N-1} n_i^2 +
	(M-\sum_{i=1}^{N-1} n_i)^2- M \right].
\end{equation}
Then, from the conditions
\begin{equation}
	\left(\frac{\partial C_T}{\partial n_i}\right)_{n_{j \neq i}}=0,\;\;\; \mbox{for}\;\; i=1,...,N-1,
\end{equation}
we find
\begin{equation}
	n_i=M-\sum_{j=1}^{N-1} n_j \;\;\; \mbox{for}\;\; i=1,...,N-1,
	\label{min-conditions}
\end{equation}
which leads to the result stated in equation (\ref{min_config}).

Thus, for a given number $N$ of orders (or springs), the minimum of $C_T$ is
\begin{equation}
	C_{T,0}(N,M)=N C_O + \frac{C_H}{2}M\left(\frac{M}{N}-1 \right),
	\label{minct}
\end{equation}
which reaches the minimum for
\begin{equation}
	N_{\tt op}=\left\{
	\begin{array}{lll}
		M   & \mbox{if} & \gamma \le 1 \\
		\gamma^{-1/2} M & \mbox{if} & \gamma > 1\;\; ,
	\end{array}
	\right.
	\label{func-por-partes}
\end{equation}
where
\begin{equation}
	\gamma=\frac{2C_O}{C_H}.
	\label{func-gamma}
\end{equation}
We conclude that the optimal strategy for minimizing the total cost involves placing $N_{\tt op}$ equally spaced orders. Thus, the optimal total cost results
\begin{equation}
	C_T^{\tt op}= C_{T,0}(N_{\tt op},M)= M C_O f(\gamma),
	\label{ctop}
\end{equation}
where
\begin{equation}
	f(\gamma)=\left\{
	\begin{array}{lll}
		1  & \mbox{if} & \gamma \le 1 \\
		\frac{2\sqrt {\gamma}-1}{\gamma}        & \mbox{if} & \gamma > 1\;\;
	\end{array}
	\right.
	\label{func-partes}
\end{equation}
represents the ratio between the optimal total cost and the total cost without storage, i.e., $C_T^{\tt op}/(M C_O$).

Equations (\ref{func-por-partes})-(\ref{func-partes}) indicate that when the quantities required in each period are equal, the optimal solution will be given by $N_{\tt op}$ orders of equal size ($n_i$=$M/N_{\tt op}$). If the quotient $M/N_{op}$ is an integer, all $n_i$ values are equal. However, if the remainder $R$ of the quotient  $M/N_{op}$ is nonzero ($R \neq 0$), the optimal solution consists of a combination of two different $n_i$ values. Specifically, the solution includes $(N_{op} -R)$ orders with $n_i$ equal to the integer part of $M/N_{op}$ and $R$ orders with $n_i$ equal to the integer part of $M/N_{op}$ plus one. This point will be discussed in more detail in the next section.

In summary, this section has shown that the lot size problem can be significantly simplified, not only in its mathematical formulation but also in the understanding of its optimization (minimization) conditions, when analyzed through its analogy with a system composed of an array of particles connected by linear springs. The equations developed here, along with the analysis to be conducted in the next section, will provide a simple and practical theoretical framework, enabling agents or organizations to make quick and reliable decisions during the supply process. 

\section{Results and discussion}
\label{results}

In the first part of this section (subsection \ref{primerparte}), we analyze the results obtained from the equations presented in the previous section, which are used to determine the optimal lot size in a production system. The second part of the section (subsection \ref{realcase}) is dedicated to the analysis of a real case study involving a concentrated pulp processing plant.  

\subsection{Theoretical analysis of the model presented in the previous section } \label{primerparte}

To understand the fundamental phenomenology, we first examine the case of a fixed planning horizon $M$ and varying values of the parameter $\gamma$.

\begin{figure}
\begin{center}
	\includegraphics[width=0.65\linewidth]{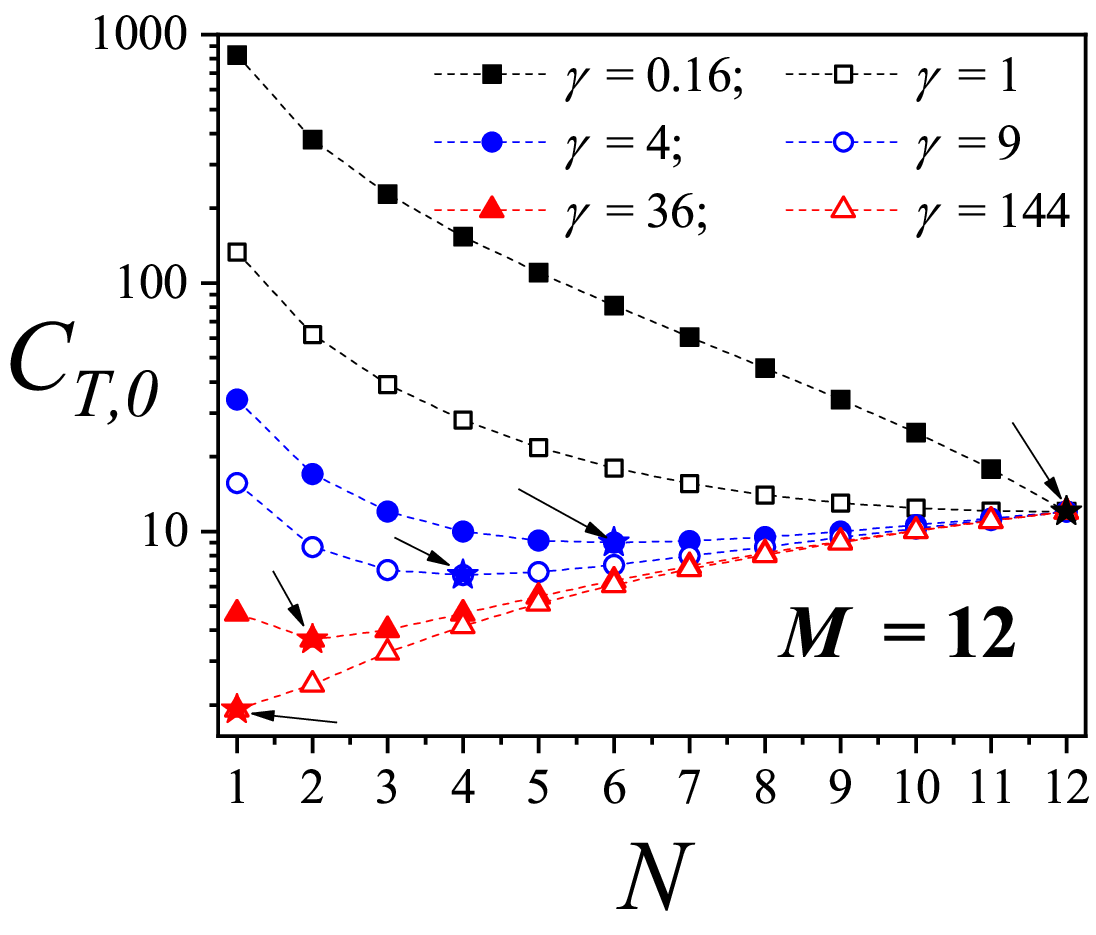}
	\caption{The minimum total cost $C_{T,0}$ (in $\log$ scale) as a function of the number of orders $N$ for a planning horizon of $M = 12$, with $C_O=1.0$ and different values of $\gamma$. The various values of $\gamma$ were achieved by adjusting $C_H$: $C_H=25/2$ ($\gamma = 0.16$); $C_H=2.0$ ($\gamma = 1.0$); $C_H=1/2$ ($\gamma = 4.0$); $C_H=2/9$ ($\gamma = 9.0$); $C_H=1/18$ ($\gamma = 36$) and $C_H=1/72$ ($\gamma = 144$).}
	\label{fig3}
\end{center}
\end{figure}

Figure \ref{fig3} illustrates the minimum total cost $C_{T,0}$ as a function of the number of orders $N$ for a planning horizon of $M=12$. Different curves correspond to distinct values of $\gamma$: 0.16 (solid squares), 1 (open squares), 4 (solid circles), 9 (open circles), 36 (solid triangles), and 144 (open triangles). In all cases, the order cost $C_O$ is set to unity, while the holding cost $C_H$ is adjusted to achieve the desired values of $\gamma$. A logarithmic scale is used on the $y$-axis to enhance the visualization of the curves topology.

For low values of $\gamma$, where $C_H \gg C_O$, it is more cost-effective to increase the number of orders and maintain lower inventory levels. Consequently, the total cost reaches a minimum when $N=N_{\tt op}=M$ (equivalent to placing an order in each period). As indicated by Eq. \ref{func-por-partes}, this scenario occurs for $0 < \gamma \leq 1$. In Fig. \ref{fig3}, this condition is observed for $\gamma=0.16$ (solid squares) and $\gamma=1$ (open squares), where the minimum total cost is achieved at $N_{\tt op}=12$. Note that the minimum point on each curve is marked with a star symbol and highlighted with an arrow.

At the other extreme, for high values of $\gamma$, where $C_H \ll C_O$, the total cost is minimized by purchasing all the material at the start of the work plan ($N = 1$). This scenario is depicted in Fig. \ref{fig3} for $\gamma=144$ (open triangles). According to Eq. \ref{func-por-partes}, the optimal number of orders is given by 
$=N_{\tt op} = \gamma^{-1/2} M =1$.

Figure \ref{fig3} also depicts several intermediate scenarios: $\gamma=4$ with $N_{\tt op} =6$  (solid circles); $\gamma=9$  with $N_{\tt op} =4$ (open circles); and $\gamma=36$  with $N_{\tt op} =2$ (solid triangles). In all instances, the number of orders is less than the number of periods in the planning horizon ($N_{\tt op} < M$). Consequently, given the higher value of the ordering cost relative to the holding cost, it becomes advantageous to reduce the number of orders by holding additional raw materials in inventory. 

In Fig. \ref{fig3}, and for the sake of clarity, the values of $\gamma$ were chosen such that the quotient $M/N_{\tt op}$ is an integer and all $n_i$ values are equal. Let us now examine what happens when this is not the case. Specifically, we analyze the scenario where $\gamma = 2.0$. For this value of $\gamma$, the relationship between ordering cost  ($C_O$) and holding cost ($C_H$) is $C_O/C_H = 1.0$ [see Eq. (\ref{func-gamma})]. Furthermore, the number of orders that minimizes the total cost is $M/(\sqrt 2)=12.0/(\sqrt 2)\approx 8.485$ [see Eq. (\ref{func-por-partes})]. This value is rounded to the nearest whole number, yielding $N_{\tt op} = 8$ (as only a whole number of orders can be made in a real production system). If the system does not permit fractional values of $n_i$, the optimal solution involves two distinct values for $n_i$. Let us now describe the procedure to determine these values.

We begin by analyzing the quotient $M/N_{\tt op}$. In this case $M = 12$ and $N_{\tt op} = 8$. Therefore,
\begin{equation}
	\qquad \frac {M}{N_{\tt op}}=\frac{12}{8}=1 R 4. \nonumber 
\end{equation}
After obtaining the integer part of $M/N_{\tt op}$ (1 in this case) and the remainder 
$R$ (4 in this case), the optimal solution can be expressed in terms of these quantities. Specifically, the minimum total cost is achieved by combining 
$(N_{\tt op} -R)$ orders with $n_i$ equal to the integer part of $M/N_{\tt op}$, and $R$ orders with $n_i$ equal to the integer part of $M/N_{\tt op}+1$. In this particular case, where $\gamma = 2.0$, the optimal solution consists of four orders with $n_i=1$ and four orders with $n_i=2$.


\begin{figure}
\begin{center}
	\includegraphics[width=0.65\linewidth]{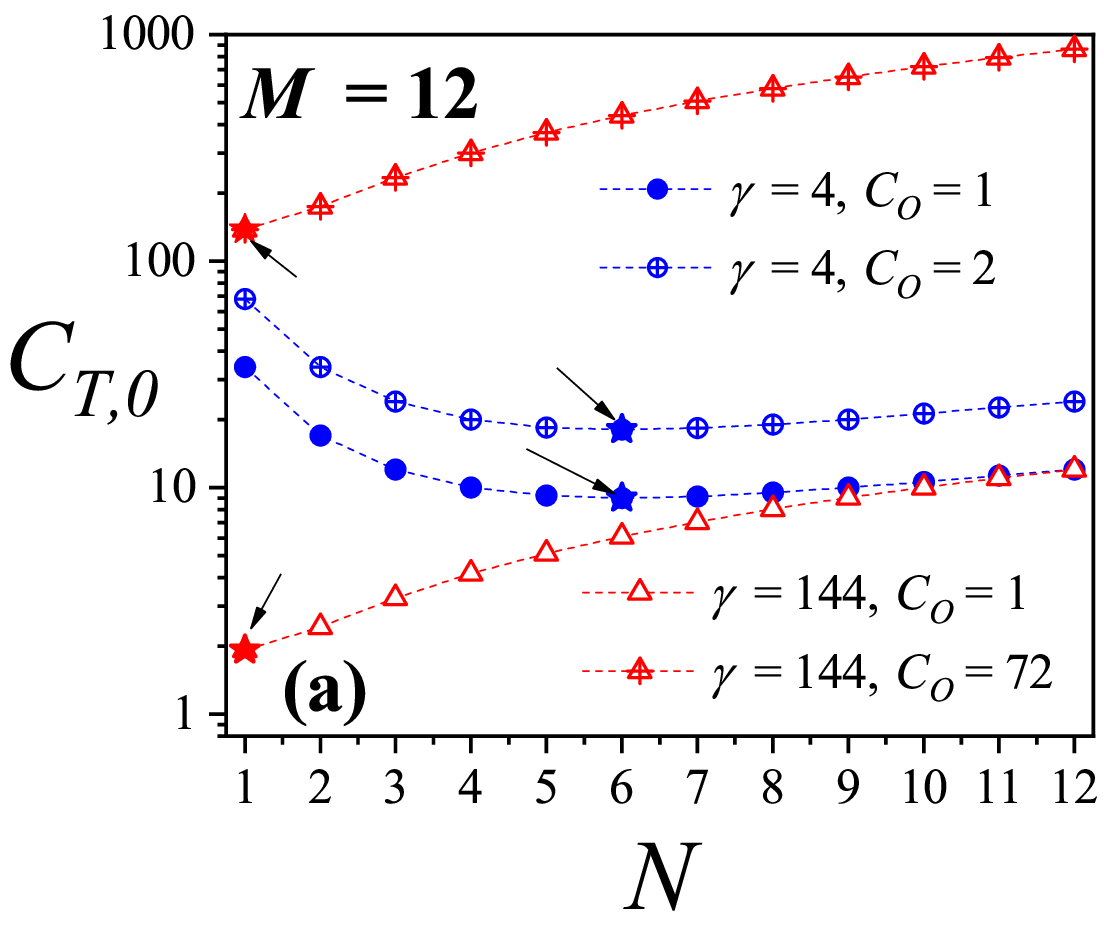}
	\includegraphics[width=0.65\linewidth]{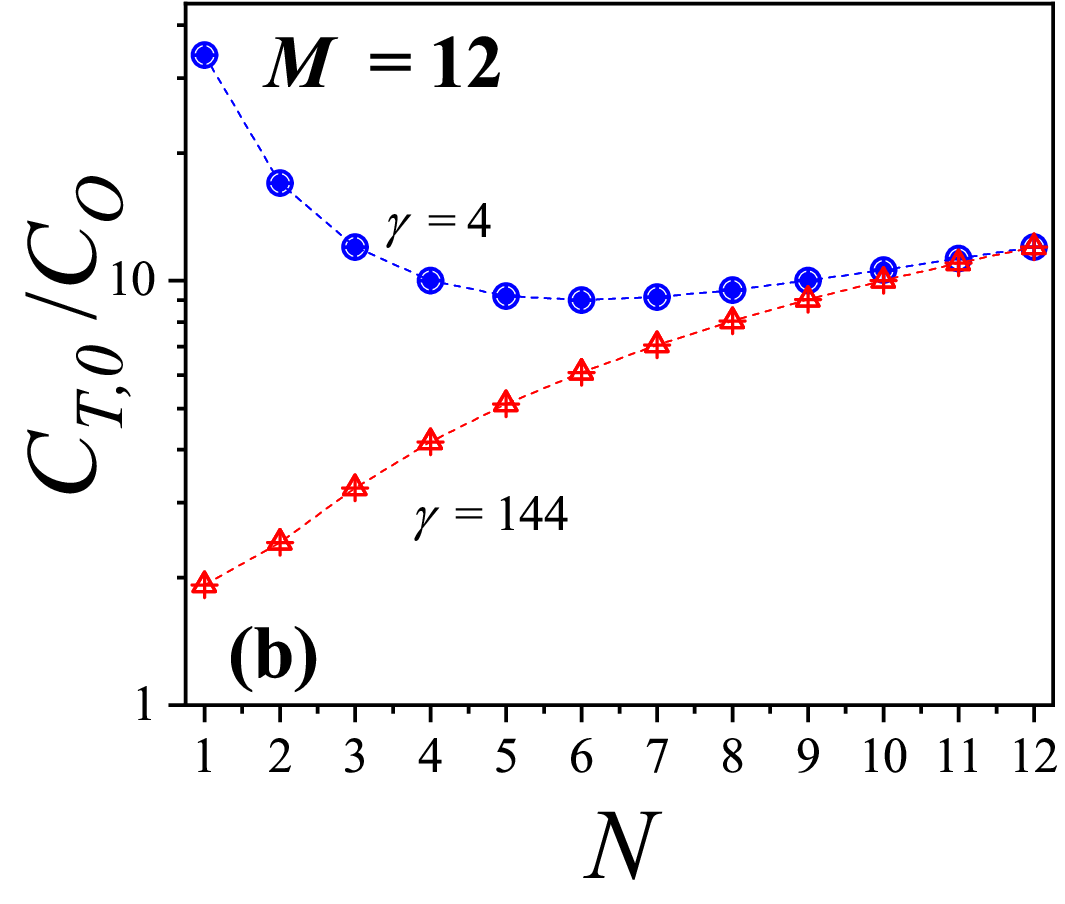}
	\caption{(a) The minimum total cost $C_{T,0}$ (in $\log$ scale) versus amount of orders $N$ for four typical cases: $\gamma=4$, $C_O=1$ and $C_H=1/2$, solid circles; $\gamma=4$, $C_O=2$ and $C_H=1$, circles with a plus inside; $\gamma=144$, $C_O=1$ and $C_H=1/72$, open triangles; and $\gamma=144$, $C_O=72$ and $C_H=1$, triangles with a plus inside. All curves were obtained for $M=12$. In part (b), the curves presented in part (a) have been normalized by the ordering cost $C_O$. With this rescaling, it is observed that all curves corresponding to the same values of $\gamma$ and $M$ collapse into a single curve. }
	\label{fig4}
\end{center}
\end{figure}

We now proceed to study the effect of jointly varying the ordering cost and the holding cost while keeping the parameter $\gamma$ and the planning horizon $M$ constant. Two representative cases are selected for comparison: one corresponding to 
$\gamma=4$, $M=12$ and another to $\gamma=144$, $M=12$. The results are presented in Fig. \ref{fig4}(a), where circles represent the outcomes for $\gamma=4$ ($C_O=1$ and $C_H=1/2$, solid circles; and $C_O=2$ and $C_H=1$, circles with a plus inside), while triangles correspond to the data obtained for $\gamma=144$ ($C_O=1$ and $C_H=1/72$, open triangles; and $C_O=72$ and $C_H=1$, triangles with a plus inside). As in the previous figure, a logarithmic scale is used on the $y$-axis.

As observed in Fig. \ref{fig4}(a), the shape of the curves and the minimization conditions remain unchanged for each fixed value of $\gamma$. The only effect of simultaneously varying $C_O$ and $C_H$ (while keeping $\gamma$ and $M$ fixed) is a vertical shift of the $C_{T,0}$ curve along the vertical axis. This behavior becomes more evident when rescaling the $C_{T,0}$ curves by $C_O$ ($C_{T,0}/C_O$). Namely, when the minimum total cost is normalized by the ordering cost, all curves corresponding to the same values of $\gamma$ and $M$ collapse into a single curve. This is illustrated in Fig. \ref{fig4}(b) for the curves shown in part (a).


\begin{figure}
\begin{center}
	\includegraphics[width=0.65\linewidth]{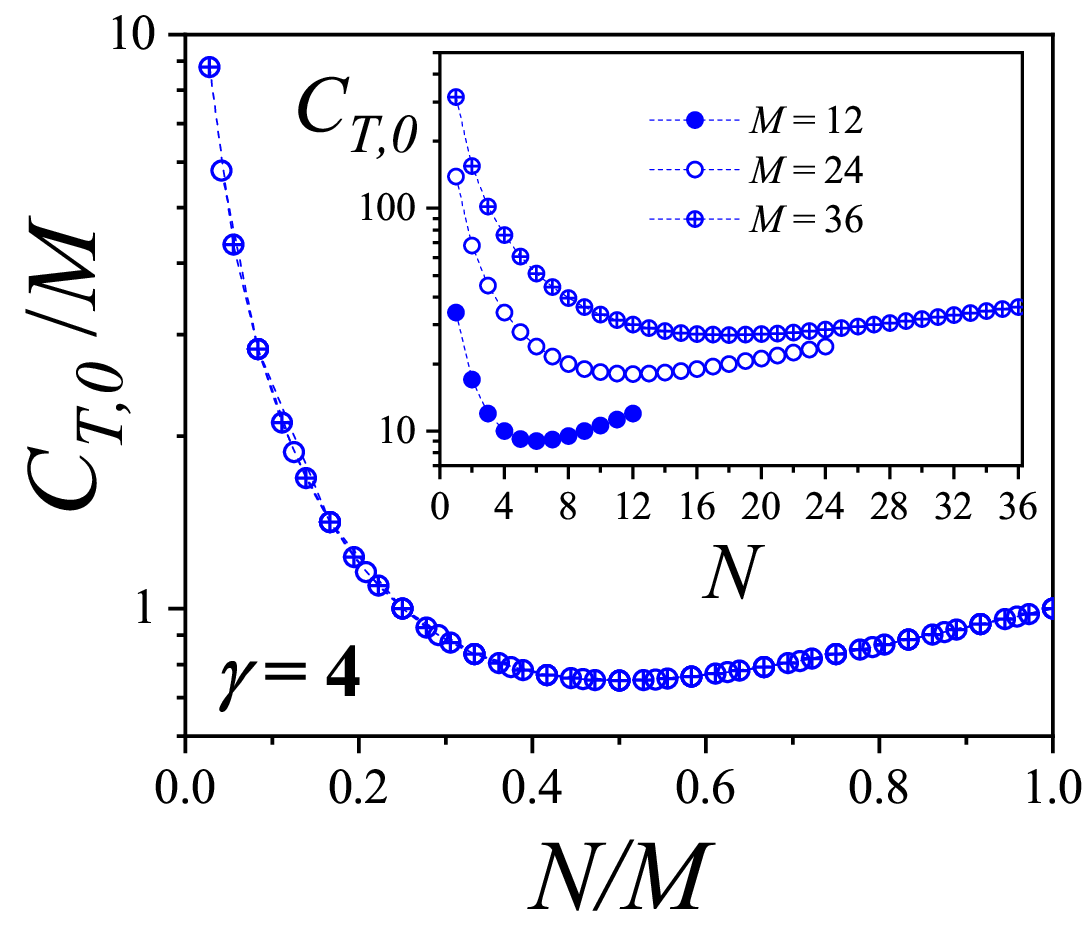}
	\caption{Data collapse of the minimum total cost $C_{T,0}/M$ (in $\log$ scale) versus $N/M$ for the case $\gamma = 4$. The plots were made using $C_O=1.0$, $C_H = 1/2$ and three different values of $M$: $M=12$, solid circles; $M=24$, open circles; and $M=36$, circles with a plus inside. The non-normalized curves ($C_{T,0}$ versus $N$) corresponding to the data in the main figure are displayed in the inset.}
	\label{fig5}
\end{center}
\end{figure}

In the previous Figs. \ref{fig3} and \ref{fig4}, the calculations were performed for a fixed planning horizon of $M=12$. In Fig. \ref{fig5}, the effect of varying $M$ is analyzed. To this end, the inset of Fig. \ref{fig5} presents the minimum total cost as a function of $N$ for a fixed value of the parameter $\gamma$ (specifically, 
$\gamma=4$ with $C_O=1.0$ and $C_H=1/2$) and three different values of $M$: $M=12$ (solid circles), $M=24$ (open circles), and $M=36$ (circles with a plus inside).

Since $\gamma=4$, all three curves exhibit a minimum at $N=M/\sqrt{4}=M/2$. Additionally, the curves shift upward as $M$ increases. When both axes are normalized by dividing by $M$, the curves corresponding to different values of $M$ collapse into a single curve, as shown in the main part of Fig. \ref{fig5}. The collapses observed in Figs. \ref{fig4}(b) and \ref{fig5} demonstrate the robustness of using $\gamma$ as a control parameter for the system. This behavior will become even more evident in the analysis of the next figure.


\begin{figure}
\begin{center}
	\includegraphics[width=0.65\linewidth]{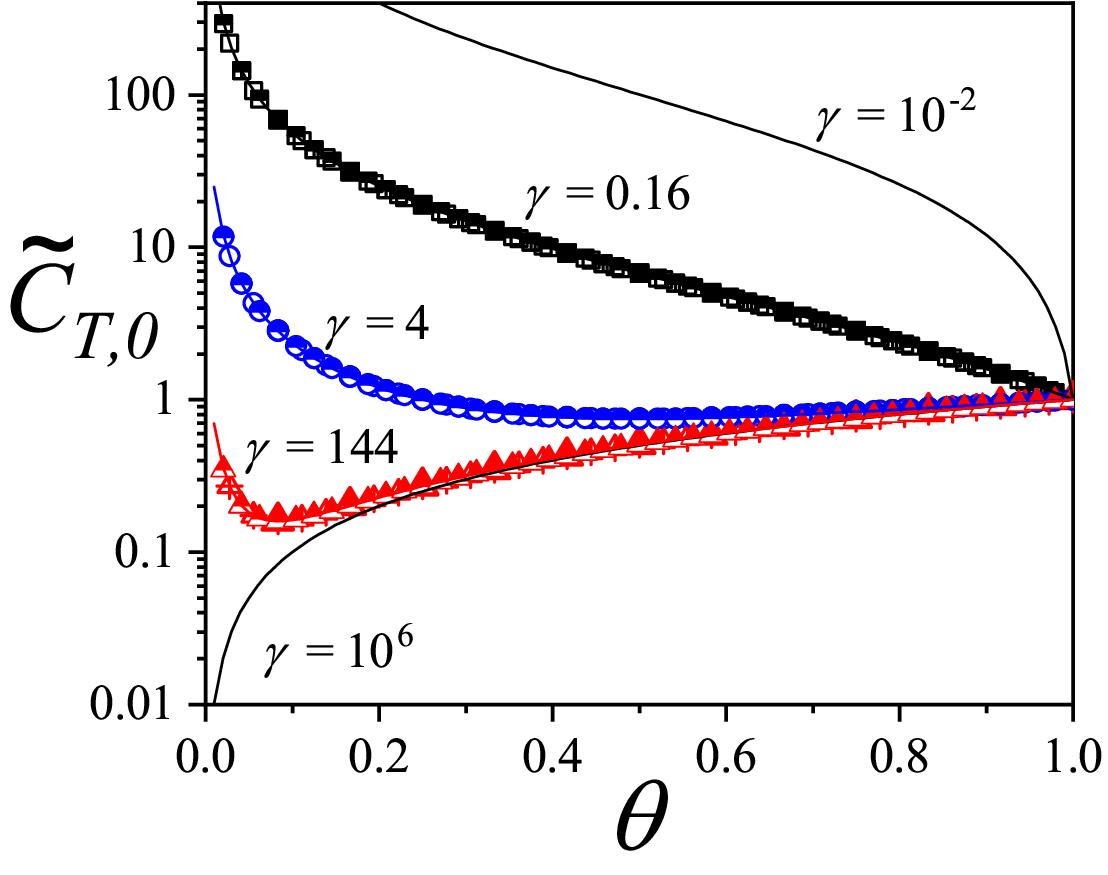}
	\caption{$\widetilde{C}_{T,0}(\theta)$ as a function of $\theta$ for different values of the parameter $\gamma$: $10^{-2}$, 0.16, 4, 144 and $10^6$, as indicated. Solid lines represent the results obtained from equation \ref{minctita}. Symbols correspond to different sets of parameters 
			$C_O-C_H-M$ that are compatible with the same $\gamma$ value.   $\gamma=0.16$: $C_O=0.08$, $C_H=1$, $M=12$ (solid squares); $C_O=0.16$, $C_H=2$, $M=24$ (squares with a plus inside); $C_O=0.32$, $C_H=4$, $M=36$ (open squares); and $C_O=0.48$, $C_H=6$, $M=48$ (squares with a point inside). $\gamma=4$: $C_O=2$, $C_H=1$, $M=12$ (solid circles); $C_O=4$, $C_H=2$, $M=24$ (circles with a plus inside); $C_O=8$, $C_H=4$, $M=36$ (open circles); and $C_O=12$, $C_H=6$, $M=48$ (circles with a point inside). 
			$\gamma=144$: $C_O=72$, $C_H=1$, $M=12$ (open triangles); $C_O=144$, $C_H=2$, $M=24$ (solid triangles); $C_O=288$, $C_H=4$, $M=36$ (triangles with a plus inside); and $C_O=432$, $C_H=6$, $M=48$ (triangles with a point inside).   }
	\label{fig6} 
\end{center}
\end{figure}

The analysis presented in Figs. \ref{fig4} and \ref{fig5} suggests a strategy for constructing the minimum cost function in terms of intensive variables. This approach provides a means to condense the entire phenomenology of the problem into a single function. In practical terms, given the ordering and holding costs, as well as the desired planning horizon, all relevant information can be encapsulated in a single function, allowing the agent to make purchasing decisions faster and more accurately. To achieve this, we simply divide equation \ref{minct} by $MC_O$ and express the resulting equation in terms of the intensive variable $\theta=N/M$, as follows:
\begin{equation}
	\frac{C_{T,0}(N,M)}{MC_O}=\frac{N}{M} + \frac{C_H}{2C_O}\left(\frac{M}{N}-1 \right),
	\label{minctnot}
\end{equation}
and  
\begin{equation}
	\widetilde{C}_{T,0}(\theta)=\theta + \frac{1}{\gamma}\left(\frac{1}{\theta}-1 \right),
	\label{minctita}
\end{equation}
where $\widetilde{C}_{T,0}(\theta)=C_{T,0}(N,M)/(MC_O)$.

In Fig. \ref{fig6}, the behavior of $\widetilde{C}_{T,0}(\theta)$ as a function of 
$\theta$ is presented for five different values of the parameter $\gamma$, as indicated in the figure. For illustrative purposes and to highlight the behavior of 
$\widetilde{C}_{T,0}(\theta)$ in the limits of low and high $\gamma$ values, two extreme cases are included: $\gamma=10^{-2}$ and $\gamma=10^6$. These cases define the upper and lower bounds of the curves shown in the figure.

For intermediate values of $\gamma$ (0.16, 4, and 144), each curve includes results for four different sets of parameters $C_O-C_H-M$ (see the figure caption). This demonstrates that equation \ref{minctita} follows a superscaling law, encapsulating all the necessary information in terms of a single parameter, $\gamma$. This property makes equation \ref{minctita} a valuable tool for organizations to optimize their logistics management.

In the next section \ref{realcase}, the framework developed here will be applied to validate the technique and demonstrate its versatility in analyzing real data from an actual processing plant.

\subsection{Application to real supply data from a concentrated pulp processing plant} \label{realcase}

In this section, we analyze a real-world purchasing and delivery process of a single product from a supplier to a customer to assess the scope and limitations of the theoretical framework discussed. The data were obtained from a concentrated pulp processing plant specializing in tomato, peach, pumpkin, and plum pulp, among others. This facility is located in the Industrial Park of San Rafael, Mendoza, Argentina.

The container used for pulp packaging, known as a polyethylene bag, is one of the company's most critical products due to its high turnover. This packaging material is the focus of the present study. The parameter values provided by the company are as follows: planing horizon $M=25$ days, ordering cost per order $C_O=477.22$ \$ and holding cost per period $C_H=9.48648$ \$. Without following any formal optimization criteria, the material was purchased through 13 orders ($N=13$) as follows: $n_1=2$; $n_2=3$ and $n_3=n_4= \dots=n_{13}=2$. The resulting total cost was 5859.49 \$. 

Next, we make a comparative cost analysis between the company's current strategy and the one proposed in this paper. Based on the data provided by the company, 
\begin{equation}
	\gamma=\frac{2C_O}{C_H}=\frac{2*477.22}{9.48648}=100.6106,
\end{equation}
\begin{equation}
	N_{\tt op}=\frac{M}{\sqrt\gamma}=\frac{25}{\sqrt{100.6106}}= 2.492,
\end{equation}
and the optimal quantity of orders is defined by the nearest integer, $N_{\tt op}=2$.
If the system allowed fractional values of $n_i$, the solution would be  $n_i=M/N_{\tt op}=12.5$ and, according to equation (\ref{ctop}), the optimal total cost would be:
\begin{equation}
	C_T^{\tt op}= M C_O \left( \frac{2\sqrt {\gamma}-1}{\gamma} \right)= 2260.27  \ \$. 
\end{equation}

However, in this study case, the values of $n_i$ must be integers. Since the quotient $M/N_{\tt op}$ is not an integer, the solution consists of a combination of two different values for $n_i$. Specifically, the solution includes $N_{\tt op} - R$ orders of size $n_i$ equal to the quotient $M/N_{\tt op}$ and $R$ orders of size $M/N_{\tt op} + 1$.
Thus,
\begin{equation}
	\qquad \frac {M}{N_{\tt op}}=\frac{25}{2}=12 R 1, \nonumber 
\end{equation}
resulting in a solution that combines one order ($N_{\tt op} - R=1$) of size 12 ($n_1=12$) and one order ($R=1$) of size 13 ($n_2=13$). The corresponding total cost results $C_{T,0}(N=1,M=12) + C_{T,0}(N=1,M=13) = 2320.49$ \$. Clearly, the purchasing strategy derived from our model significantly reduces costs compared to the strategy previously used by the company. This outcome was expected since, as mentioned earlier, the company did not make purchasing decisions based on any optimization criteria.

Knowing that the classical EOQ method \cite{Harris,Caliskan} is valid for determining the size of order lots when demand is constant, we will use it to compare with the proposed model. 

Before conducting the comparison and analyzing the process from the perspective of the EOQ model, it is useful to express the ordering and holding parameters in a more appropriate form. In this context, the holding cost per period $C_H$ can be rewritten as the material requirement per period $\alpha$ multiplied by the cost of holding a single unit of the product in inventory, $c_h$: $C_H= \alpha c_h$. For the case under study, $\alpha=174$ units, and accordingly, $c_h= C_H /\alpha= 9.48648/174=0.05452$ \$.  

The EOQ formula determines the number of product units, $EOQ$, that should be ordered to minimize holding and ordering costs. This quantity is given by:
\begin{eqnarray}
	EOQ & = & \left(\frac{2 \times {\rm material \ requirement \ per \ period} \times {\rm ordering \ cost \ per \ order}}{{\rm holding \ cost \ of \  a \ single \ unit}}\right)^{1/2} \nonumber \\
	& = & \left(\frac{2 \times \alpha \times C_O}{c_h}\right)^{1/2}.
	\label{EOQ}
\end{eqnarray}
The strategy proposed by the EOQ model begins with the acquisition of $EOQ$ product units at the start of the process. Once these units are consumed, a new order of $EOQ$ product units is placed. This cycle repeats until the end of the work period.

Next, we will analyze the application of the EOQ model to the case under study, which will provide a better understanding of the EOQ strategy. 

By applying equation \ref{EOQ} with $\alpha=174$ units, $C_O=477.22$ \$ and $c_h=0.05452$ \$, the EOQ value is obtained as $EOQ=1745.3$ units. It is straightforward to see that $n_1$ can be calculated as $n_1=EOQ/\alpha$. In this case, $EOQ/\alpha=10.03$. Rounding to ensure $n_1$ is an integer gives $n_1=10$. This first order will be consumed over the first ten days, requiring a new order of $n_2=10$ on day 11. The second order will be used between days 11 and 20. Consequently, a new order must be placed on day 21; however, the planning horizon $M=25$ days constrains the size of the third order to $n_3=5$. With this strategy, consisting of two orders of size 10 and one order of size 5, the resulting total cost is 2380.31 \$, slightly higher than the cost obtained using the model proposed in this study. 

\begin{table}[h]
\begin{center}
	\begin{tabular}{|c|c|c|c|}
		\hline
		& \# of orders    & Set $\{n_i\}$    & Total cost \\ \hline
		Company strategy  & 13 & $n_1=2$, $n_2=3$, $n_3=n_4= \dots=n_{13}=2$ & 5859.49 \$  \\ \hline
		EOQ model & 3 & $n_1=10$, $n_2=10$, $n_3=5$ & 2380.31 \$  \\ 		\hline
		Our model & 2 & $n_1=12$, $n_2=13$ & 2320.49 \$  \\ 		\hline
	\end{tabular}
	\caption{\label{tabla_1} Cost parameter values used by the company for procuring polyethylene bags, along with those obtained from the application of the EOQ model and our model for the case under study.}
\end{center}
\end{table}


\begin{figure}
\begin{center}        
    \includegraphics[width=0.65\linewidth]{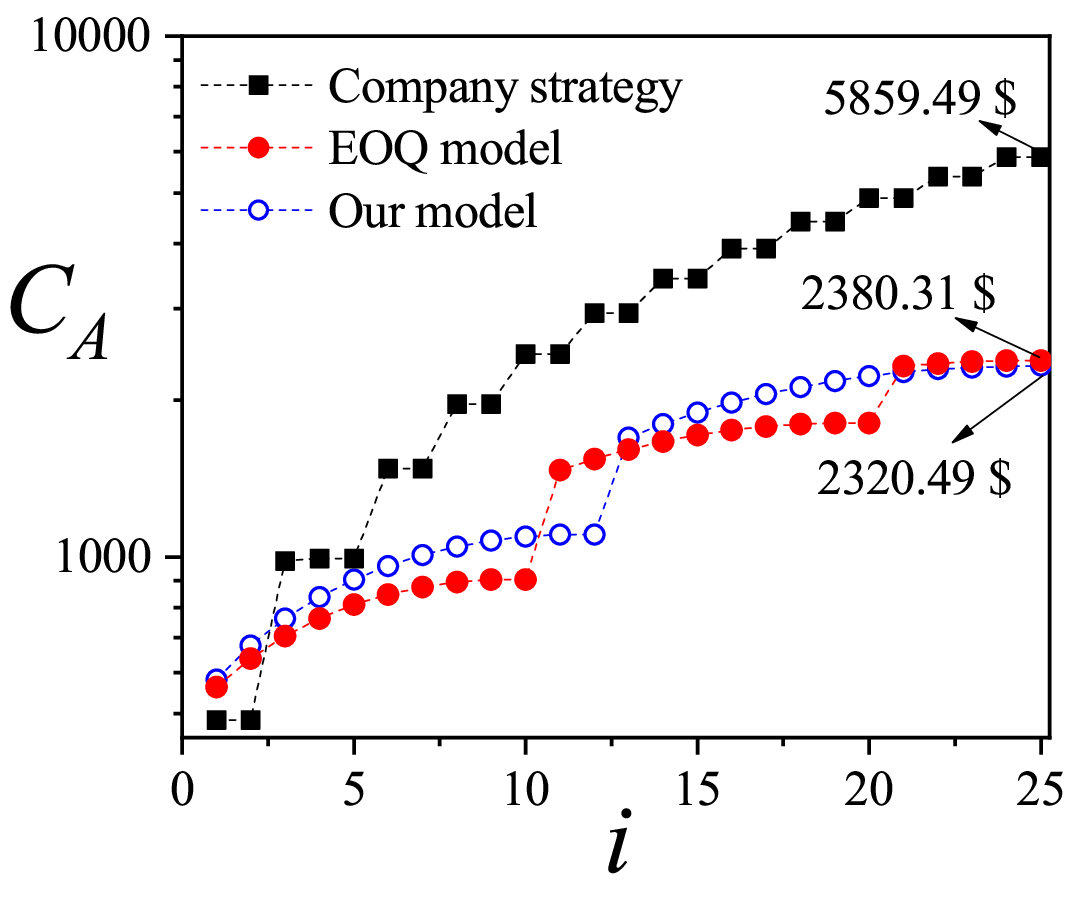}
	\caption{Accumulated total cost $C_A$ as a function of the number of days $i$, according to the EOQ method (solid circles), our model (open circles), and the company's strategy (solid squares).}
	\label{fig7}
\end{center}
\end{figure}

The cost parameters characterizing the three purchasing strategies discussed in this section are compiled in Table \ref{tabla_1}. Additionally, these strategies can be easily visualized through the total accumulated cost, $C_A(i)$, which represents the sum of the ordering and holding costs on day $i$. Figure \ref{fig7} illustrates $C_A$ versus $i$ for the EOQ model (solid circles), the strategy chosen by the company (solid squares), and our model (open circles). Both the EOQ model and our model show very similar behavior, with a small difference observed in the total cost at the end of the cycle. On the other hand, the strategy used by the company appears highly inefficient, highlighting the importance of adopting an appropriate purchasing strategy to improve production costs.

To conclude this study, it is worth noting that the execution time required to solve the case under investigation in this section was analyzed using (i) the proposed model, (ii) the All Possible Combinations (APC) model previously developed by our group \cite{Tania2023}, and (iii) several widely accepted models from the literature: Fixed Order Quantity  \cite{FOQ}, Periodic Order Quantity \cite{FOQ}, Lot-for-Lot \cite{Noori1997}, Wagner-Whitin algorithm \cite{Wagner}, Silver-Meal algorithm \cite{Silver}, and Mixed Integer Linear Programming \cite{Akimoto,Golmohamadi}. The results obtained indicate that the execution time of the proposed model is considerably lower than that of the other models used in the comparison.
\\

\section{Conclusions}
\label{conclu}

In this work, we have presented an optimization model for material lot sizing using an elastic system. We have established that the problem of material supply is isomorphic to a simple one-dimensional mechanical system of point particles connected by elastic elements. By leveraging this isomorphism with a well-known physical system, the cost optimization conditions emerge naturally, and the exact solution to the problem can be readily obtained. 

The optimal number of orders, $N_{\tt op}$, and their respective sizes, $n_i$ ($i=1, \dots, N_{\tt op}$), were determined in terms of a parameter $\gamma$, which relates the ordering cost per order $C_O$ and the holding cost per period for the quantity of material required in one period $C_H$, such that $\gamma=2C_O/C_H$. This parameter $\gamma$ fully governs the system’s behavior. 

If $\gamma \leq 1$ (i.e., the ordering cost is less than or equal to half the product of the holding cost and material procurement), the optimal number of orders equals the system size, $N_{\tt op} = M$. This implies that the optimal purchasing policy consists of placing one order per period, $n_i=1$.

For $\gamma > 1$, the optimal number of orders is expected to be lower than the planning horizon $M$. Our analytical result, $N_{\tt op}=\gamma^{-1}M$, confirms this expectation and enables the formulation of a new material supply policy under limited ordering. In this scenario, two cases arise: (i) if the quotient $M/N_{\tt op}$ is an integer, then 
$N_{\tt op}$ orders of equal size are placed, with $n_i$=$M/N_{\tt op}=\sqrt{\gamma}$;
and (ii) if the remainder $R$ of the quotient $M/N_{\tt op}$ is nonzero, then 
$(N_{\tt op} -R)$ orders are placed with $n_i$ equal to the integer part of $M/N_{\tt op}$, and the remaining $R$ orders have $n_i$ equal to the integer part of 
$M/N_{\tt op}+1$.

By expressing the minimum total cost function in terms of the intensive variable 
$N/M$, we demonstrate that the proposed approach condenses the entire phenomenology of the problem into a single function depending solely on $\gamma$. This allows us to solve the lot-sizing problem through a single equation, without relying on algorithms or extensive computations—one of the key contributions of this work. In practical terms, given the ordering and holding costs, as well as the desired planning horizon, all relevant information is encapsulated within a single function, enabling faster and more accurate purchasing decisions.

The model was also applied to a real-world case and compared with well-established classical algorithms. The solution provided by our model not only achieves excellent cost optimization but also exhibits the shortest execution time.

In summary, this study introduces new insights into the implementation of optimization methods for material lot sizing using an elastic system. The results of this preliminary study are promising, and the proposed theoretical model can be readily extended to incorporate additional complexities. Future research will focus on adapting the model to scenarios with variable material demands, quantity discounts, and deteriorating items. Work in this direction is currently underway.

\section{Acknowledgements}
This work was supported in part by CONICET (Argentina)  under Project No. PIP 11220220100238CO; Universidad Nacional de San Luis (Argentina) under Project 03-1920; Universidad Nacional de Mar  del Plata (Argentina) under project EXA-989/20; and Universidad Tecnol\'ogica  Nacional, Facultad Regional San Rafael, under projects PID UTN PAECBSR0008105TC and PATCSR0010050TC.

\end{document}